\documentclass[amsmath,amssymb,reprint,aps,prl, floatfix]{revtex4-1}
\usepackage{bm}
\usepackage{graphicx}
\usepackage{braket}
\usepackage{epstopdf}
\usepackage{bm}
\usepackage{lineno}
\usepackage{natbib}

\begin{document}

\title{Determining complementary properties with quantum clones}

\author{G.S. Thekkadath{*}}

\author{R.Y. Saaltink}

\author{L. Giner}

\author{J.S. Lundeen}

\affiliation{Department of Physics, Centre for Research in Photonics, University
of Ottawa, 25 Templeton Street, Ottawa, Ontario K1N 6N5, Canada~\\
 {*}gthek044@uottawa.ca}
\begin{abstract}
In a classical world, simultaneous measurements of complementary properties
(\textit{e.g.} position and momentum) give a system's state. In quantum
mechanics, measurement-induced disturbance is largest for complementary properties
and, hence, limits the precision with which such properties can be
determined simultaneously. It is tempting to try to sidestep this
disturbance by copying the system and measuring each complementary
property on a separate copy. However, perfect copying is physically
impossible in quantum mechanics. Here, we investigate using the closest quantum analog to this copying strategy, optimal cloning. The coherent portion of the generated clones' state corresponds to ``twins'' of the input system. Like perfect copies, both twins faithfully reproduce the properties of the input system. Unlike perfect copies, the twins are entangled. As such, a measurement on both twins is equivalent to a simultaneous measurement on the input system. For complementary observables, this joint measurement gives the system's state, just as in the classical case. We demonstrate this experimentally using polarized single photons.
\end{abstract}
\maketitle

At the heart of quantum mechanics is the concept of complementarity:
the impossibility of precisely determining complementary properties
of a single quantum system. For example, a precise measurement of
the position of an electron causes a subsequent momentum measurement
to give a random result. Such joint measurements are the crux of Heisenberg's
measurement-disturbance relation~\cite{ozawa2004uncertainty,busch2014colloquium}, as highlighted
by his famous microscope thought-experiment in 1927~\cite{heisenberg1927quantum}.
Since then, methods for performing joint measurements of complementary
properties have been steadily theoretically investigated~\cite{wigner1932on,arthurs1965simultaneous,muynck1979simultaneous,braginsky1980quantum,carmeli2012informationally},
leading to seminal inventions such as heterodyne quantum state tomography~\cite{shapiro1984phase,leonhardt1993phase}. More recently, advances in the ability to
control measurement-induced disturbance have led to
ultra-precise measurements that surpass standard quantum limits~\cite{colangelo2017simultaneous},
and also simultaneous determination of complementary properties with
a precision that saturates Heisenberg's bound~\cite{ringbauer2014joint}. In sum, joint complementary measurements continue to prove useful for
characterizing quantum systems~\cite{lundeen2011direct,thekkadath2016direct,bamber2014observing,salvail2013full} and for understanding foundational issues in quantum mechanics~\cite{ringbauer2014joint,gourgy2016quantum,colangelo2017simultaneous,rozema2012violation}.

In this Letter, we address the main challenge in performing a joint
measurement, which is to circumvent the mutual disturbance caused
by measuring two general non-commuting observables, $\bm{X}$ and
$\bm{Y}$. Classically, such joint measurements (\textit{e.g.} momentum and
position) are sufficient to determine the state of the system, even of statistical
ensembles. In quantum mechanics, these joint measurements have mainly
been realized by carefully designing them to minimize their disturbance,
such as in weak~\cite{lundeen2011direct,thekkadath2016direct,bamber2014observing,salvail2013full,rozema2012violation,ringbauer2014joint}
or non-demolition~\cite{braginsky1980quantum,gourgy2016quantum,colangelo2017simultaneous}
measurements. In order to avoid these technically complicated measurements, one might instead consider manipulating the system,
and in particular, copying it. Subsequently, one would perform a standard
measurement separately on each copy of the system. Since the measurements
are no longer sequential, or potentially not in the same location,
one would not expect them to physically disturb one another. Crucially, as we explain below, the copies being measured must be
correlated for this strategy to work. Hofmann recently proposed an experimental procedure that achieves this~\cite{hofmann2012how}. Following his proposal, we experimentally demonstrate that a partial-SWAP two-photon quantum logic gate~\cite{cernoch2008experimental} can
isolate the measurement results of two photonic ``twins". These twins are quantum-correlated (\textit{i.e.} entangled) copies of a photon's polarization state that are ideal for performing joint measurements.

We begin by considering a physically impossible, but informative,
strategy. Given a quantum system in a state $\bm{\rho}$, consider
making two perfect copies $\bm{\rho}\otimes\bm{\rho}$ and then measuring
observable $\bm{X}$ on copy one and $\bm{Y}$ on copy two. In this
case, the joint probability of measuring outcomes $X=x$ and $Y=y$
is $\mathrm{Prob}(x,y)=\mathrm{Prob}(x)\mathrm{Prob}(y)$~\footnote{As is
usual for a probability, $\mathrm{Prob}(x,y)$ is estimated from repeated trials using an identical ensemble of input systems. This is implicit for probabilities and expectation values throughout the paper.}. Since it is factorable into functions
of $x$ and $y$, this joint probability cannot reveal correlations
between the two properties. Even classically, this procedure would generally
fail to give the system's state, since such correlations can occur
in \textit{e.g.} statistical ensembles. Less obviously, these correlations can occur
in a single quantum system due to quantum coherence~\cite{wigner1932on}.
In turn, the lack of sensitivity to this coherence makes this joint
measurement informationally incomplete~\cite{muynck1979simultaneous},
and thus this simplistic strategy is insufficient for determining
quantum states~\footnote{The strategy considered here is informationally equivalent to separating an identical ensemble into two and measuring $\mathrm{Prob}(x)$ with one half and $\mathrm{Prob}(y)$ with the other half. Knowing only these two marginal distributions is insufficient to determine the quantum state~\cite{lvovsky2009continuous}.}. Further confounding this strategy, the no-cloning theorem
prohibits any operation that can create a perfect copy of an arbitrary
quantum state, $\bm{\rho}\nrightarrow\bm{\rho}\otimes\bm{\rho}$~\cite{wootters1982single}.
In summary, even if this strategy were allowed in quantum physics,
it would not function well as a joint measurement.

Although perfect quantum copying is impossible, there has been extensive
work investigating ``cloners" that produce imperfect copies~\cite{scarani2005quantum}.
Throughout this paper, we consider a general ``$1\rightarrow2$ cloner".
It takes as an input an unknown qubit state $\bm{\rho}_{a}$ along
with a blank ancilla $\bm{I}_{b}/2$ ($\bm{I}$ is the identity operator),
and attempts to output two copies of $\bm{\rho}$ into separate modes,
$a$ and $b$.

We now consider a second strategy, one that utilizes a trivial version
of this cloner by merely shuffling the modes of the two input states.
This can be achieved by swapping their modes half of the time, and
for the other half, leaving them unchanged. That is, one applies with
equal likelihood the SWAP operation ($\bm{S}_{ab}$: $\bm{\rho}_{a}\bm{I}_{b}/2\rightarrow\bm{I}_{a}\bm{\rho}_{b}/2$),
or the identity operation ($\bm{I}_{ab}=\bm{I}_{a}\otimes\bm{I}_{b}$):
\begin{equation}
\bm{\rho}_{a}\bm{I}_{b}/2\rightarrow(\bm{\rho}_{a}\bm{I}_{b}+\bm{I}_{a}\bm{\rho}_{b})/4\equiv\bm{t}_{ab}.\label{eqn:trivial_clone}
\end{equation}
Each output mode of the trivial cloner $\bm{t}_{ab}$ contains an
imperfect copy of the input state $\bm{\rho}$. Jointly measuring
$\bm{X}$ and $\bm{Y}$, one on each trivial clone, yields the result
$\mathrm{Prob}(x,y)=(\mathrm{Prob}(x)+\mathrm{Prob}(y))/4$. In contrast
to a joint measurement on perfect copies, this result exhibits correlations
between $x$ and $y$. These appear because in any given trial, only
one of the observables is measured on $\bm{\rho}$, while the other
is measured on the blank ancilla. Hence, the apparent correlations
are an artifact caused by randomly switching the observable being
measured, and are not due to genuine correlations that could be present
in $\bm{\rho}$. While now physically allowed, this joint measurement
strategy is still insufficient to determine the quantum state $\bm{\rho}$. 

In order to access correlations in the quantum state, we must take
advantage of quantum coherence. Instead of randomly applying $\bm{S}_{ab}$
or $\bm{I}_{ab}$ as in trivial cloning, we require the superposition
of these two processes, \textit{i.e.} the coherent sum: 
\begin{equation}
\bm{\Pi}_{ab}^{j}=\frac{1}{2}(\bm{I}_{ab}+j\bm{S}_{ab}),\label{eqn:general_sym}
\end{equation}
where now we are free to choose the phase $j$. $\bm{\Pi}^{j}$ is
a generalized symmetry operation that can implement a partial-SWAP
gate~\cite{cernoch2008experimental}. For $j=+1$ ($-1$), this
operation is a projection onto the symmetric (anti-symmetric) part
of the trivial cloner input, $\bm{\rho}_{a}\bm{I}_{b}/2$. The symmetric subspace only contains
states that are unchanged by a SWAP operation. A projection onto this
subspace increases the relative probability that $\bm{\rho}_{a}$
and the blank ancilla are identical. In fact, it has been proven that
a symmetric projection on the trivial cloner input
is the optimal cloning process, since it maximizes the fidelity of
the clones (\textit{i.e.} their similarity to $\bm{\rho}$)~\cite{buzek1996quantum,gisin1997optimal,bruss1998optimal}.

This brings us to our third and final strategy. Optimal cloning achieves
more than just producing imperfect copies: the clones are quantum-correlated,
\textit{i.e.} entangled~\cite{buzek1996quantum}. This can be seen
by examining the output state of the optimal cloner (i.e. with $j=1)$:
\begin{equation}
\bm{o}_{ab}^{j}=\frac{2}{3}(\bm{\Pi}_{ab}^{j}\bm{\rho}_{a}\bm{I}_{b}\bm{\Pi}_{ab}^{j\dagger})=\frac{2}{3}\bm{t}_{ab}+\frac{1}{3}\mathrm{Re}\left[j\bm{c}_{ab}\right],\label{eqn:opt_clone2}
\end{equation}
where $\bm{c}_{ab}=\bm{S}_{ab}\bm{\rho}_{a}\bm{I}_{b}$ and $\mathrm{Re}\left[\mathbf{s}\right]=(\mathbf{s}+\mathbf{s}^{\dagger})/2$.
While the first term is two trivial clones, the second term is the
coherent portion of the optimal clones, and is the source of their
entanglement. Considered alone, $\bm{c}_{ab}$ corresponds to two
``twins" of $\bm{\rho}$. Like perfect copies, any measurement on either twin gives results identical to what would be obtained with
$\bm{\rho}$~\cite{hofmann2012how}. However, the twins are entangled. As such, it is important
to realize that they are very different from the uncorrelated perfect
copies we considered in the first strategy. Relative to these (\textit{i.e.} $\bm{\rho}\otimes\bm{\rho}$), performing the same joint measurement as before, but
on the twins $\bm{c}_{ab}$, provides more information about $\bm{\rho}$. Measuring
$\bm{X}$ on one twin and $\bm{Y}$ on the other yields the expectation
value $\braket{\bm{xy}}_{\bm{\rho}}=\mathrm{Tr}(\bm{xy\rho})$, where
$\bm{x}=\ket{x}\bra{x}$ and $\bm{y}=\ket{y}\bra{y}$ are projectors
onto the eigenstates of observables $\bm{X}$ and $\bm{Y}$, respectively.
Classically, this result would be interpreted as a joint probability
$\mathrm{Prob}(x,y)$. However,
due to Heisenberg's uncertainty principle, $\braket{\bm{xy}}_{\bm{\rho}}$
has non-classical features that shield precise determination of both
$\bm{X}$ and $\bm{Y}$. In fact, $\braket{\bm{xy}}_{\bm{\rho}}$ is a ``quasiprobability"
distribution much like the Wigner distribution~\cite{wigner1932on}, and has similar properties such as being rigorously equivalent to the state $\bm{\rho}$~\cite{bamber2014observing}.
Unlike the Wigner distribution, it is generally complex
since $\bm{xy}$ is not an observable (\textit{i.e.} it is non-Hermitian).
Although the measurements of $\bm{X}$ and $\bm{Y}$ are performed
independently on each twin, because the twins are entangled, it is
equivalent to simultaneously measuring the same two observables on
a single copy of $\bm{\rho}$. This approach is complementary to other joint measurement strategies for state determination in which the measurement itself is entangling, while the copies being measured are separable~\cite{niset2007superiority,massar1995optimal}.

Performing a joint measurement directly on twins cannot be achieved
in a physical process. This is likely part of the reason why previous theoretical investigations concluded that optimal cloners were not ideal for joint measurements~\cite{buzek1996quantum,dariano2001joint,brougham2006cloning}. However, in a joint measurement on optimal
clones, Hofmann showed that the contribution from the twins can be isolated from that
of the trivial clones~\cite{hofmann2012how}. This is because changing the phase $j$ affects
only the coherent part of the cloning process. Thus, by adding joint
measurement results obtained from the optimal cloner with different
phases $j$, we can isolate the contribution from the twins and measure
$\braket{\bm{xy}}_{\bm{\rho}}$~\cite{SM}.

\begin{figure}
\centering \includegraphics[width=0.25\textwidth]{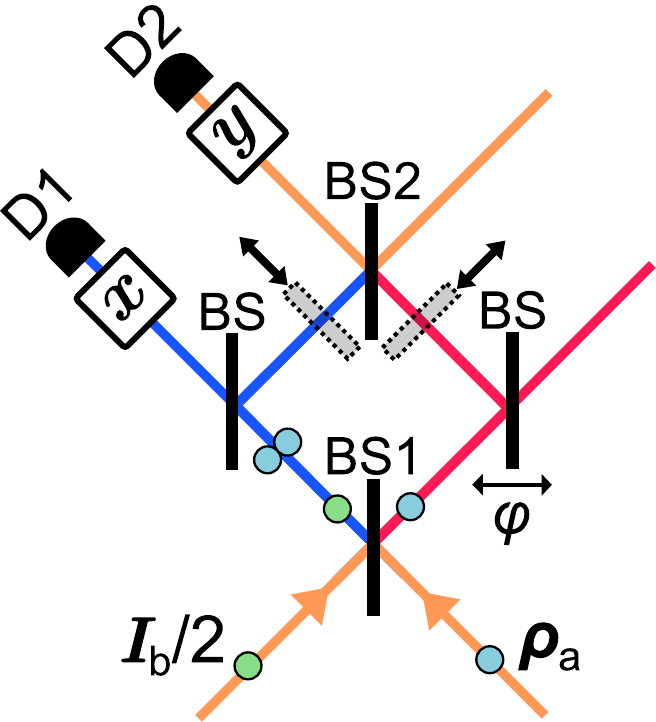} \caption{\textbf{Schematic of experimental setup.} A photon in a polarization
state $\bm{\rho}_{a}$ and a photon in a blank state $\bm{I}_{b}/2$
enter an interferometer containing removable beam blocks (dotted outline).
Complementary observables $\bm{x}$ and $\bm{y}$ are jointly measured
by counting coincidences at detectors D1 and D2. When the red (blue)
path is blocked, we post-select on the case where the photons exit
the first beam splitter BS1 from the same (opposite) port and perform
a symmetric projector $\bm{\Pi}^{+1}$ (anti-symmetric projector $\bm{\Pi}^{-1}$),
thus making two optimal clones of $\bm{\rho}$. With no path blocked
and a phase difference of $\varphi=\pm\pi/2$ between paths, we coherently
combine both cases and perform $\bm{\Pi}^{\pm i}$, respectively.}
\label{fig:exp_scheme} 
\end{figure}

The experiment is shown schematically in Fig.~\ref{fig:exp_scheme}.
A photonic system lends itself to optimal cloning because the symmetry
operation $\bm{\Pi}^{j}$ in Eq.~\ref{eqn:general_sym} can be implemented
with a beam splitter (BS). If two indistinguishable photons impinge
onto different ports of BS1, Hong-Ou-Mandel interference occurs and
the photons always ``bunch" by exiting BS1 from a single
port. By selecting cases where photons bunch (anti-bunch),
one implements the symmetry projector $\bm{\Pi}^{+1}$ ($\bm{\Pi}^{-1}$)~\cite{campos2005permutation}.
This enabled previous experimental demonstrations of optimal cloners
for both polarization~\cite{irvine2004optimal} and orbital angular
momentum~\cite{nagali2009optimal,bouchard2016high} states. However,
we must also implement $\bm{\Pi}^{\pm i}$. Following a similar strategy
as Refs.~\cite{cernoch2008experimental,maclean2016quantum}, we
use an interferometer to coherently combine the symmetric and anti-symmetric
projectors, since $\bm{\Pi}^{\pm i}=(e^{\pm i\pi/4}\bm{\Pi}^{+1}+e^{\mp i\pi/4}\bm{\Pi}^{-1})/\sqrt{2}$.
This is achieved by interfering at BS2 the cases where the photons
bunched at BS1 with cases where they anti-bunched at BS1. In summary, this
provides an experimental procedure to vary the phase $j$ and thereby
isolate the joint measurement contribution of the twins from that
of the trivial clones.

\begin{figure}
\centering \includegraphics[width=0.43\textwidth]{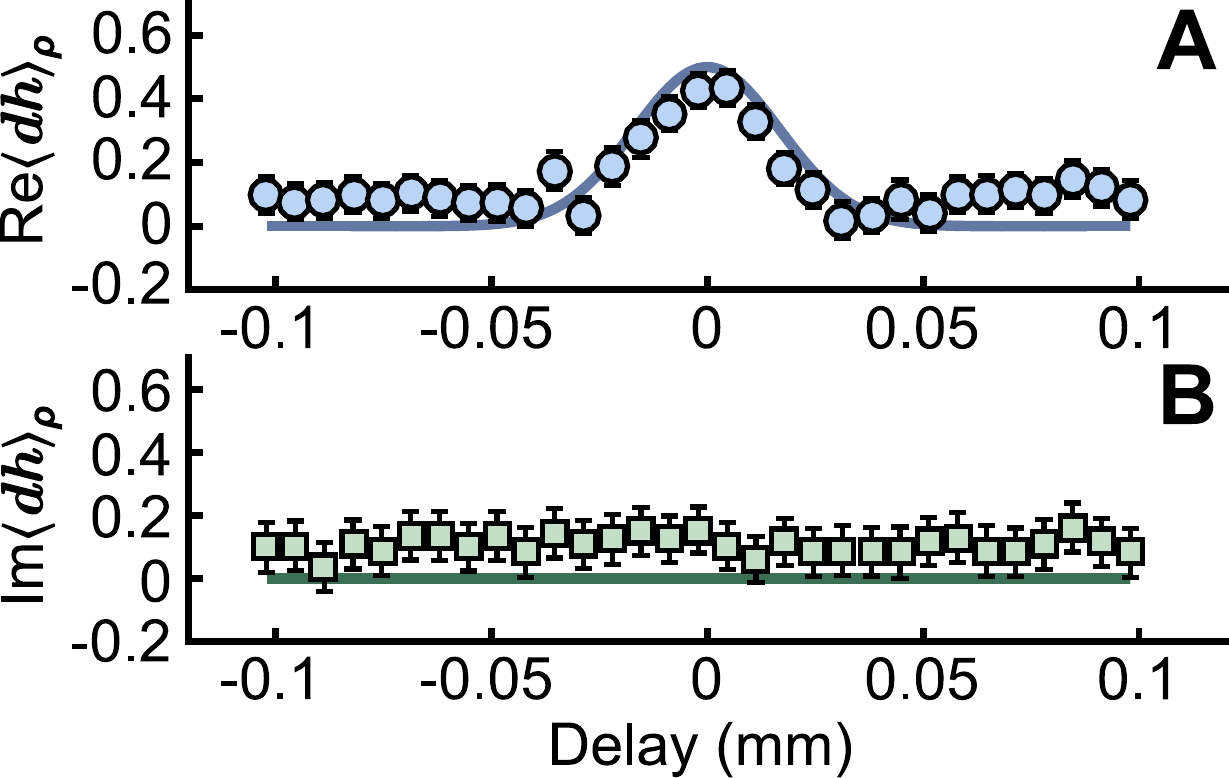}
\caption{\textbf{Transition from trivial to optimal cloning.} A horizontal
photon $\bm{\rho}_{a}=\bm{h}$ is sent into the cloner. We jointly
measure complementary observables $\bm{d}$ and $\bm{h}$, one on
each clone, and plot the real (\textbf{A}) and imaginary (\textbf{B})
parts of $\braket{\bm{dh}}_{\bm{\rho}}$. For large delays, only trivial
clones are produced. Since they contain no information about $\braket{\bm{dh}}_{\bm{\rho}}$,
our procedure cancels their contribution to the joint measurement
result. At zero delay, optimal clones are produced. We isolate the
contribution of the twins to the joint measurement, yielding the desired
value of $\braket{\bm{dh}}_{\bm{\rho}}=0.5$. The bold lines are theory
curves calculated for intermediate delays~\cite{SM}. Error bars
are calculated using Poissonian counting statistics.}
\label{fig:stage} 
\end{figure}

We experimentally verify that this procedure works by performing a
joint measurement on trivial clones $\bm{t}_{ab}$ and showing that
its outcome does not contribute to $\braket{\bm{xy}}_{\bm{\rho}}$.
In particular, we scan the delay between $\bm{\rho}_{a}$ and $\bm{I}_{b}/2$
at BS1. When the delay is zero, we implement the symmetry operator
$\bm{\Pi}^{j}$. When the delay is larger than the coherence time
of the photons, the BS does not discriminate the symmetry of the two-qubit
input state. Thus, it simply shuffles the modes of both qubits and
produces trivial clones $\bm{t}_{ab}$. We test the procedure by measuring
$\braket{\bm{xy}}_{\bm{\rho}}=\braket{\bm{dh}}_{\bm{\rho}}$, where
$\bm{d}$ and $\bm{h}$ are diagonal and horizontal polarization projectors,
respectively. We use an input state $\bm{\rho}_{a}=\bm{h}$, for which
one expects $\braket{\bm{dh}}_{\bm{\rho}}=\mathrm{Tr}(\bm{dhh})=0.5$.
In Fig.~\ref{fig:stage}, we show that for large delays $\braket{\bm{dh}}_{\bm{\rho}}=0$,
whereas for zero delay, it obtains its full value. This shows that
the procedure has effectively removed the contribution of the trivial
clones to the optimal clone state in Eq. \ref{eqn:opt_clone2}, and so the joint measurement result is solely due to the twins.

\begin{figure*}
\centering 
\includegraphics[width=0.8\textwidth]{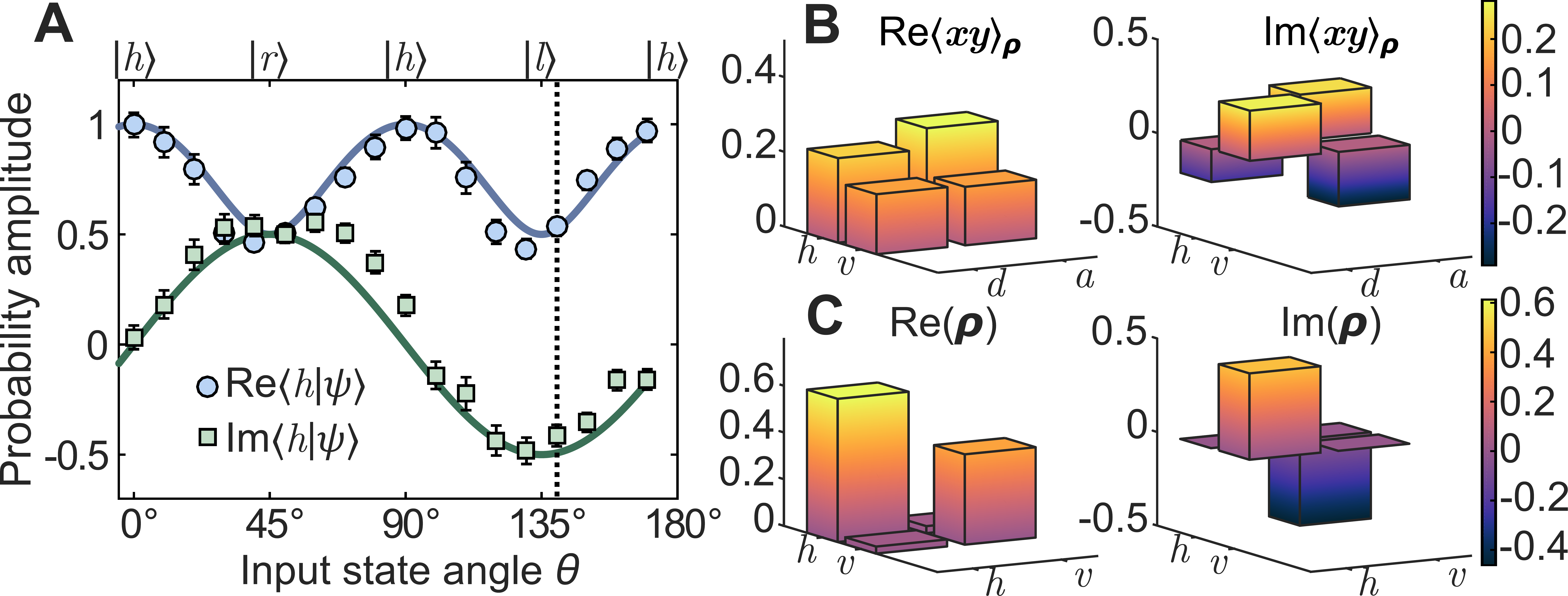}
\caption{\textbf{Measuring the quantum state.} Various polarization states
$\ket{\psi}=\alpha\ket{h}+\beta\ket{v}$ are produced by rotating
the fast-axis angle $\theta$ of a quarter-wave plate with increments
of $10^{\circ}$. We plot the real and imaginary parts of $\alpha=\braket{h|\psi}=\sqrt{\frac{3}{8}\cos(4\theta)+\frac{5}{8}}+i\sin{\theta}\cos{\theta}$
in \textbf{A} (theory is bold lines, $\ket{r}=(\ket{v}+i\ket{h})/\sqrt{2}$
and $\ket{l}=(\ket{v}-i\ket{h})/\sqrt{2}$ are circular polarizations).
Error bars are calculated using Poissonian counting statistics. The
entire joint quasiprobability distribution (\textbf{B}) and density
matrix (\textbf{C}) are also shown for the input state indicated by
the dashed line (color represents amplitude). After processing the
counts with a maximum-likelihood estimation, the average fidelity
$\left|\braket{\psi|\bm{\rho}|\psi}\right|^{2}$ of the 18 measured
states is $0.92\pm0.05$.}
\label{fig:densitystates} 
\end{figure*}

A joint measurement on twins of $\bm{\rho}$ can reveal correlations
between complementary properties in $\bm{\rho}$. We measure the entire
joint quasiprobability distribution $\braket{\bm{xy}}_{\bm{\rho}}$
for the complementary polarization observables $\bm{x}=\{\bm{d},\bm{a}\}$
using diagonal and anti-diagonal projectors, and $\bm{y}=\{\bm{h},\bm{v}\}$
using horizontal and vertical projectors. This is repeated for a variety
of different input states $\bm{\rho}$. For the input state indicated
by the dashed line in Fig.~\ref{fig:densitystates}A, correlations
can be seen in $\mathrm{Im}\braket{\bm{xy}}_{\bm{\rho}}$, as shown
in Fig.~\ref{fig:densitystates}B. With the ability to exhibit correlations,
$\braket{\bm{xy}}_{\bm{\rho}}$ is now a complete description of the
quantum state $\bm{\rho}$~\cite{SM}. In particular, the wave function
of the state (see Fig.~\ref{fig:densitystates}A) is any cross-section
of $\braket{\bm{xy}}_{\bm{\rho}}$. Moreover, the density matrix (see
Fig.~\ref{fig:densitystates}C) can be obtained with a Fourier transform
of $\braket{\bm{xy}}_{\bm{\rho}}$. This is the key experimental result.
In the classical world, simultaneously measuring complementary properties gives the system's state. This result demonstrates that simultaneously measuring complementary observables on twins, similarly, gives the system's state.

In addition to its fundamental importance, our result has potential practical advantages as a state determination
procedure. It is valid
for higher dimensional states~\cite{SM} for which standard quantum tomography
requires prohibitively many measurements. Specifically, a $d$-dimensional state typically requires $\mathcal{O}(d^2)$ measurements in $\mathcal{O}(d)$ bases to be reconstructed tomographically. In contrast, here
the wave function is obtained directly (\textit{i.e.}
without a reconstruction algorithm) from $4d$ experimental measurements of only two observables, $\bm{X}$ and $\bm{Y}$.

Our results uncover striking connections with other joint measurement
techniques, despite the physics of each approach being substantially
different. For example, the joint quasiprobability $\braket{\bm{xy}}_{\bm{\rho}}$
is also the average outcome of another joint measurement strategy:
the weak measurement of $\bm{y}$ followed by a measurement of $\bm{x}$
on a single system $\bm{\rho}$~\cite{salvail2013full,bamber2014observing,hofmann2012how}.
Furthermore, in the continuous-variable analogue of our work, measurements of complementary observables on cloned Gaussian states~\cite{andersen2005unconditional}
give a different, but related, quasiprobability distribution for the
quantum state known as the Q-function~\cite{leonhardt1993phase}. Finally, the result of a joint measurement on phase-conjugated Gaussian states can be used in a feedforward to produce optimal clones~\cite{sabuncu2007experimental}. These connections emphasize the central role of optimal cloning in quantum
mechanics~\cite{wootters1982single,bruss1998optimal} and clarify
the intimate relation between joint measurements of complementary
observables and determining quantum states~\cite{wigner1932on,muynck1979simultaneous}.


We anticipate that simultaneous measurements of non-commuting observables
can be naturally implemented in quantum computers using our technique,
since the operation $\bm{\Pi}^{j}$ can be achieved using a controlled-SWAP
quantum logic gate~\cite{hofmann2012how,patel2016quantum}. As joint
measurements are pivotal in quantum mechanics, this will have broad
implications for state estimation~\cite{lundeen2011direct,thekkadath2016direct,bamber2014observing,salvail2013full},
quantum control~\cite{gourgy2016quantum}, and quantum foundations~\cite{rozema2012violation,ringbauer2014joint}.
For instance, we anticipate that our method can be used to efficiently
and directly measure high-dimensional quantum states that are needed
for fault-tolerant quantum computing and quantum cryptography~\cite{bouchard2016high}.

\begin{acknowledgments}
This work was supported by the Canada Research Chairs (CRC) Program,
the Natural Sciences and Engineering Research Council (NSERC), and
Excellence Research Chairs (CERC) Program.
\end{acknowledgments}

\nocite{durt2010mutually}

\bibliographystyle{apsrev4-1}
\bibliography{refs}

\begin{thebibliography}{42}%
\makeatletter
\providecommand \@ifxundefined [1]{%
 \@ifx{#1\undefined}
}%
\providecommand \@ifnum [1]{%
 \ifnum #1\expandafter \@firstoftwo
 \else \expandafter \@secondoftwo
 \fi
}%
\providecommand \@ifx [1]{%
 \ifx #1\expandafter \@firstoftwo
 \else \expandafter \@secondoftwo
 \fi
}%
\providecommand \natexlab [1]{#1}%
\providecommand \enquote  [1]{``#1''}%
\providecommand \bibnamefont  [1]{#1}%
\providecommand \bibfnamefont [1]{#1}%
\providecommand \citenamefont [1]{#1}%
\providecommand \href@noop [0]{\@secondoftwo}%
\providecommand \href [0]{\begingroup \@sanitize@url \@href}%
\providecommand \@href[1]{\@@startlink{#1}\@@href}%
\providecommand \@@href[1]{\endgroup#1\@@endlink}%
\providecommand \@sanitize@url [0]{\catcode `\\12\catcode `\$12\catcode
  `\&12\catcode `\#12\catcode `\^12\catcode `\_12\catcode `\%12\relax}%
\providecommand \@@startlink[1]{}%
\providecommand \@@endlink[0]{}%
\providecommand \url  [0]{\begingroup\@sanitize@url \@url }%
\providecommand \@url [1]{\endgroup\@href {#1}{\urlprefix }}%
\providecommand \urlprefix  [0]{URL }%
\providecommand \Eprint [0]{\href }%
\providecommand \doibase [0]{http://dx.doi.org/}%
\providecommand \selectlanguage [0]{\@gobble}%
\providecommand \bibinfo  [0]{\@secondoftwo}%
\providecommand \bibfield  [0]{\@secondoftwo}%
\providecommand \translation [1]{[#1]}%
\providecommand \BibitemOpen [0]{}%
\providecommand \bibitemStop [0]{}%
\providecommand \bibitemNoStop [0]{.\EOS\space}%
\providecommand \EOS [0]{\spacefactor3000\relax}%
\providecommand \BibitemShut  [1]{\csname bibitem#1\endcsname}%
\let\auto@bib@innerbib\@empty
\bibitem [{\citenamefont {Ozawa}(2004)}]{ozawa2004uncertainty}%
  \BibitemOpen
  \bibfield  {author} {\bibinfo {author} {\bibfnamefont {M.}~\bibnamefont
  {Ozawa}},\ }\href {\doibase https://doi.org/10.1016/j.physleta.2003.12.001}
  {\bibfield  {journal} {\bibinfo  {journal} {Phys. Lett. A}\ }\textbf
  {\bibinfo {volume} {320}},\ \bibinfo {pages} {367 } (\bibinfo {year}
  {2004})}\BibitemShut {NoStop}%
\bibitem [{\citenamefont {Busch}\ \emph {et~al.}(2014)\citenamefont {Busch},
  \citenamefont {Lahti},\ and\ \citenamefont {Werner}}]{busch2014colloquium}%
  \BibitemOpen
  \bibfield  {author} {\bibinfo {author} {\bibfnamefont {P.}~\bibnamefont
  {Busch}}, \bibinfo {author} {\bibfnamefont {P.}~\bibnamefont {Lahti}}, \ and\
  \bibinfo {author} {\bibfnamefont {R.~F.}\ \bibnamefont {Werner}},\ }\href
  {\doibase 10.1103/RevModPhys.86.1261} {\bibfield  {journal} {\bibinfo
  {journal} {Rev. Mod. Phys.}\ }\textbf {\bibinfo {volume} {86}},\ \bibinfo
  {pages} {1261} (\bibinfo {year} {2014})}\BibitemShut {NoStop}%
\bibitem [{\citenamefont {Heisenberg}(1927)}]{heisenberg1927quantum}%
  \BibitemOpen
  \bibfield  {author} {\bibinfo {author} {\bibfnamefont {W.}~\bibnamefont
  {Heisenberg}},\ }\href {\doibase 10.1007/BF01397280} {\bibfield  {journal}
  {\bibinfo  {journal} {Z. Phys.}\ }\textbf {\bibinfo {volume} {43}},\ \bibinfo
  {pages} {172} (\bibinfo {year} {1927})}\BibitemShut {NoStop}%
\bibitem [{\citenamefont {Wigner}(1932)}]{wigner1932on}%
  \BibitemOpen
  \bibfield  {author} {\bibinfo {author} {\bibfnamefont {E.}~\bibnamefont
  {Wigner}},\ }\href {\doibase 10.1103/PhysRev.40.749} {\bibfield  {journal}
  {\bibinfo  {journal} {Phys. Rev.}\ }\textbf {\bibinfo {volume} {40}},\
  \bibinfo {pages} {749} (\bibinfo {year} {1932})}\BibitemShut {NoStop}%
\bibitem [{\citenamefont {Arthurs}\ and\ \citenamefont
  {Kelly}(1965)}]{arthurs1965simultaneous}%
  \BibitemOpen
  \bibfield  {author} {\bibinfo {author} {\bibfnamefont {E.}~\bibnamefont
  {Arthurs}}\ and\ \bibinfo {author} {\bibfnamefont {J.~L.}\ \bibnamefont
  {Kelly}},\ }\href {\doibase 10.1002/j.1538-7305.1965.tb01684.x} {\bibfield
  {journal} {\bibinfo  {journal} {Bell Syst. Tech. J.}\ }\textbf {\bibinfo
  {volume} {44}},\ \bibinfo {pages} {725} (\bibinfo {year} {1965})}\BibitemShut
  {NoStop}%
\bibitem [{\citenamefont {de~Muynck}\ \emph {et~al.}(1979)\citenamefont
  {de~Muynck}, \citenamefont {Janssen},\ and\ \citenamefont
  {Santman}}]{muynck1979simultaneous}%
  \BibitemOpen
  \bibfield  {author} {\bibinfo {author} {\bibfnamefont {W.~M.}\ \bibnamefont
  {de~Muynck}}, \bibinfo {author} {\bibfnamefont {P.~A. E.~M.}\ \bibnamefont
  {Janssen}}, \ and\ \bibinfo {author} {\bibfnamefont {A.}~\bibnamefont
  {Santman}},\ }\href {\doibase 10.1007/BF00715052} {\bibfield  {journal}
  {\bibinfo  {journal} {Found. Phys.}\ }\textbf {\bibinfo {volume} {9}},\
  \bibinfo {pages} {71} (\bibinfo {year} {1979})}\BibitemShut {NoStop}%
\bibitem [{\citenamefont {Braginsky}\ \emph {et~al.}(1980)\citenamefont
  {Braginsky}, \citenamefont {Vorontsov},\ and\ \citenamefont
  {Thorne}}]{braginsky1980quantum}%
  \BibitemOpen
  \bibfield  {author} {\bibinfo {author} {\bibfnamefont {V.~B.}\ \bibnamefont
  {Braginsky}}, \bibinfo {author} {\bibfnamefont {Y.~I.}\ \bibnamefont
  {Vorontsov}}, \ and\ \bibinfo {author} {\bibfnamefont {K.~S.}\ \bibnamefont
  {Thorne}},\ }\href {\doibase 10.1126/science.209.4456.547} {\bibfield
  {journal} {\bibinfo  {journal} {Science}\ }\textbf {\bibinfo {volume}
  {209}},\ \bibinfo {pages} {547} (\bibinfo {year} {1980})}\BibitemShut
  {NoStop}%
\bibitem [{\citenamefont {Carmeli}\ \emph {et~al.}(2012)\citenamefont
  {Carmeli}, \citenamefont {Heinosaari},\ and\ \citenamefont
  {Toigo}}]{carmeli2012informationally}%
  \BibitemOpen
  \bibfield  {author} {\bibinfo {author} {\bibfnamefont {C.}~\bibnamefont
  {Carmeli}}, \bibinfo {author} {\bibfnamefont {T.}~\bibnamefont {Heinosaari}},
  \ and\ \bibinfo {author} {\bibfnamefont {A.}~\bibnamefont {Toigo}},\ }\href
  {\doibase 10.1103/PhysRevA.85.012109} {\bibfield  {journal} {\bibinfo
  {journal} {Phys. Rev. A}\ }\textbf {\bibinfo {volume} {85}},\ \bibinfo
  {pages} {012109} (\bibinfo {year} {2012})}\BibitemShut {NoStop}%
\bibitem [{\citenamefont {Shapiro}\ and\ \citenamefont
  {Wagner}(1984)}]{shapiro1984phase}%
  \BibitemOpen
  \bibfield  {author} {\bibinfo {author} {\bibfnamefont {J.}~\bibnamefont
  {Shapiro}}\ and\ \bibinfo {author} {\bibfnamefont {S.}~\bibnamefont
  {Wagner}},\ }\href {\doibase 10.1109/JQE.1984.1072470} {\bibfield  {journal}
  {\bibinfo  {journal} {IEEE J. Quant. Electron.}\ }\textbf {\bibinfo {volume}
  {20}},\ \bibinfo {pages} {803} (\bibinfo {year} {1984})}\BibitemShut
  {NoStop}%
\bibitem [{\citenamefont {Leonhardt}\ and\ \citenamefont
  {Paul}(1993)}]{leonhardt1993phase}%
  \BibitemOpen
  \bibfield  {author} {\bibinfo {author} {\bibfnamefont {U.}~\bibnamefont
  {Leonhardt}}\ and\ \bibinfo {author} {\bibfnamefont {H.}~\bibnamefont
  {Paul}},\ }\href {\doibase 10.1103/PhysRevA.47.R2460} {\bibfield  {journal}
  {\bibinfo  {journal} {Phys. Rev. A}\ }\textbf {\bibinfo {volume} {47}},\
  \bibinfo {pages} {R2460} (\bibinfo {year} {1993})}\BibitemShut {NoStop}%
\bibitem [{\citenamefont {Colangelo}\ \emph {et~al.}(2017)\citenamefont
  {Colangelo}, \citenamefont {Ciurana}, \citenamefont {Bianchet}, \citenamefont
  {Sewell},\ and\ \citenamefont {Mitchell}}]{colangelo2017simultaneous}%
  \BibitemOpen
  \bibfield  {author} {\bibinfo {author} {\bibfnamefont {G.}~\bibnamefont
  {Colangelo}}, \bibinfo {author} {\bibfnamefont {F.~M.}\ \bibnamefont
  {Ciurana}}, \bibinfo {author} {\bibfnamefont {L.~C.}\ \bibnamefont
  {Bianchet}}, \bibinfo {author} {\bibfnamefont {R.~J.}\ \bibnamefont
  {Sewell}}, \ and\ \bibinfo {author} {\bibfnamefont {M.~W.}\ \bibnamefont
  {Mitchell}},\ }\href {http://dx.doi.org/10.1038/nature21434} {\bibfield
  {journal} {\bibinfo  {journal} {Nature}\ }\textbf {\bibinfo {volume} {543}},\
  \bibinfo {pages} {525} (\bibinfo {year} {2017})}\BibitemShut {NoStop}%
\bibitem [{\citenamefont {Ringbauer}\ \emph {et~al.}(2014)\citenamefont
  {Ringbauer}, \citenamefont {Biggerstaff}, \citenamefont {Broome},
  \citenamefont {Fedrizzi}, \citenamefont {Branciard},\ and\ \citenamefont
  {White}}]{ringbauer2014joint}%
  \BibitemOpen
  \bibfield  {author} {\bibinfo {author} {\bibfnamefont {M.}~\bibnamefont
  {Ringbauer}}, \bibinfo {author} {\bibfnamefont {D.~N.}\ \bibnamefont
  {Biggerstaff}}, \bibinfo {author} {\bibfnamefont {M.~A.}\ \bibnamefont
  {Broome}}, \bibinfo {author} {\bibfnamefont {A.}~\bibnamefont {Fedrizzi}},
  \bibinfo {author} {\bibfnamefont {C.}~\bibnamefont {Branciard}}, \ and\
  \bibinfo {author} {\bibfnamefont {A.~G.}\ \bibnamefont {White}},\ }\href
  {\doibase 10.1103/PhysRevLett.112.020401} {\bibfield  {journal} {\bibinfo
  {journal} {Phys. Rev. Lett.}\ }\textbf {\bibinfo {volume} {112}},\ \bibinfo
  {pages} {020401} (\bibinfo {year} {2014})}\BibitemShut {NoStop}%
\bibitem [{\citenamefont {Lundeen}\ \emph {et~al.}(2011)\citenamefont
  {Lundeen}, \citenamefont {Sutherland}, \citenamefont {Patel}, \citenamefont
  {Stewart},\ and\ \citenamefont {Bamber}}]{lundeen2011direct}%
  \BibitemOpen
  \bibfield  {author} {\bibinfo {author} {\bibfnamefont {J.~S.}\ \bibnamefont
  {Lundeen}}, \bibinfo {author} {\bibfnamefont {B.}~\bibnamefont {Sutherland}},
  \bibinfo {author} {\bibfnamefont {A.}~\bibnamefont {Patel}}, \bibinfo
  {author} {\bibfnamefont {C.}~\bibnamefont {Stewart}}, \ and\ \bibinfo
  {author} {\bibfnamefont {C.}~\bibnamefont {Bamber}},\ }\href {\doibase
  10.1038/nature10120} {\bibfield  {journal} {\bibinfo  {journal} {Nature}\
  }\textbf {\bibinfo {volume} {474}},\ \bibinfo {pages} {188} (\bibinfo {year}
  {2011})}\BibitemShut {NoStop}%
\bibitem [{\citenamefont {Thekkadath}\ \emph {et~al.}(2016)\citenamefont
  {Thekkadath}, \citenamefont {Giner}, \citenamefont {Chalich}, \citenamefont
  {Horton}, \citenamefont {Banker},\ and\ \citenamefont
  {Lundeen}}]{thekkadath2016direct}%
  \BibitemOpen
  \bibfield  {author} {\bibinfo {author} {\bibfnamefont {G.~S.}\ \bibnamefont
  {Thekkadath}}, \bibinfo {author} {\bibfnamefont {L.}~\bibnamefont {Giner}},
  \bibinfo {author} {\bibfnamefont {Y.}~\bibnamefont {Chalich}}, \bibinfo
  {author} {\bibfnamefont {M.~J.}\ \bibnamefont {Horton}}, \bibinfo {author}
  {\bibfnamefont {J.}~\bibnamefont {Banker}}, \ and\ \bibinfo {author}
  {\bibfnamefont {J.~S.}\ \bibnamefont {Lundeen}},\ }\href {\doibase
  10.1103/PhysRevLett.117.120401} {\bibfield  {journal} {\bibinfo  {journal}
  {Phys. Rev. Lett.}\ }\textbf {\bibinfo {volume} {117}},\ \bibinfo {pages}
  {120401} (\bibinfo {year} {2016})}\BibitemShut {NoStop}%
\bibitem [{\citenamefont {Bamber}\ and\ \citenamefont
  {Lundeen}(2014)}]{bamber2014observing}%
  \BibitemOpen
  \bibfield  {author} {\bibinfo {author} {\bibfnamefont {C.}~\bibnamefont
  {Bamber}}\ and\ \bibinfo {author} {\bibfnamefont {J.~S.}\ \bibnamefont
  {Lundeen}},\ }\href {\doibase 10.1103/PhysRevLett.112.070405} {\bibfield
  {journal} {\bibinfo  {journal} {Phys. Rev. Lett.}\ }\textbf {\bibinfo
  {volume} {112}},\ \bibinfo {pages} {070405} (\bibinfo {year}
  {2014})}\BibitemShut {NoStop}%
\bibitem [{\citenamefont {Salvail}\ \emph {et~al.}(2013)\citenamefont
  {Salvail}, \citenamefont {Agnew}, \citenamefont {Johnson}, \citenamefont
  {Bolduc}, \citenamefont {Leach},\ and\ \citenamefont
  {Boyd}}]{salvail2013full}%
  \BibitemOpen
  \bibfield  {author} {\bibinfo {author} {\bibfnamefont {J.~Z.}\ \bibnamefont
  {Salvail}}, \bibinfo {author} {\bibfnamefont {M.}~\bibnamefont {Agnew}},
  \bibinfo {author} {\bibfnamefont {A.~S.}\ \bibnamefont {Johnson}}, \bibinfo
  {author} {\bibfnamefont {E.}~\bibnamefont {Bolduc}}, \bibinfo {author}
  {\bibfnamefont {J.}~\bibnamefont {Leach}}, \ and\ \bibinfo {author}
  {\bibfnamefont {R.~W.}\ \bibnamefont {Boyd}},\ }\href {\doibase
  10.1038/nphoton.2013.24} {\bibfield  {journal} {\bibinfo  {journal} {Nat.
  Photon.}\ }\textbf {\bibinfo {volume} {7}},\ \bibinfo {pages} {316} (\bibinfo
  {year} {2013})}\BibitemShut {NoStop}%
\bibitem [{\citenamefont {Hacohen-Gourgy}\ \emph {et~al.}(2016)\citenamefont
  {Hacohen-Gourgy}, \citenamefont {Martin}, \citenamefont {Flurin},
  \citenamefont {Ramasesh}, \citenamefont {Whaley},\ and\ \citenamefont
  {Siddiqi}}]{gourgy2016quantum}%
  \BibitemOpen
  \bibfield  {author} {\bibinfo {author} {\bibfnamefont {S.}~\bibnamefont
  {Hacohen-Gourgy}}, \bibinfo {author} {\bibfnamefont {L.~S.}\ \bibnamefont
  {Martin}}, \bibinfo {author} {\bibfnamefont {E.}~\bibnamefont {Flurin}},
  \bibinfo {author} {\bibfnamefont {V.~V.}\ \bibnamefont {Ramasesh}}, \bibinfo
  {author} {\bibfnamefont {K.~B.}\ \bibnamefont {Whaley}}, \ and\ \bibinfo
  {author} {\bibfnamefont {I.}~\bibnamefont {Siddiqi}},\ }\href
  {http://dx.doi.org/10.1038/nature19762} {\bibfield  {journal} {\bibinfo
  {journal} {Nature}\ }\textbf {\bibinfo {volume} {538}},\ \bibinfo {pages}
  {491} (\bibinfo {year} {2016})}\BibitemShut {NoStop}%
\bibitem [{\citenamefont {Rozema}\ \emph {et~al.}(2012)\citenamefont {Rozema},
  \citenamefont {Darabi}, \citenamefont {Mahler}, \citenamefont {Hayat},
  \citenamefont {Soudagar},\ and\ \citenamefont
  {Steinberg}}]{rozema2012violation}%
  \BibitemOpen
  \bibfield  {author} {\bibinfo {author} {\bibfnamefont {L.~A.}\ \bibnamefont
  {Rozema}}, \bibinfo {author} {\bibfnamefont {A.}~\bibnamefont {Darabi}},
  \bibinfo {author} {\bibfnamefont {D.~H.}\ \bibnamefont {Mahler}}, \bibinfo
  {author} {\bibfnamefont {A.}~\bibnamefont {Hayat}}, \bibinfo {author}
  {\bibfnamefont {Y.}~\bibnamefont {Soudagar}}, \ and\ \bibinfo {author}
  {\bibfnamefont {A.~M.}\ \bibnamefont {Steinberg}},\ }\href {\doibase
  10.1103/PhysRevLett.109.100404} {\bibfield  {journal} {\bibinfo  {journal}
  {Phys. Rev. Lett.}\ }\textbf {\bibinfo {volume} {109}},\ \bibinfo {pages}
  {100404} (\bibinfo {year} {2012})}\BibitemShut {NoStop}%
\bibitem [{\citenamefont {Hofmann}(2012)}]{hofmann2012how}%
  \BibitemOpen
  \bibfield  {author} {\bibinfo {author} {\bibfnamefont {H.~F.}\ \bibnamefont
  {Hofmann}},\ }\href {\doibase 10.1103/PhysRevLett.109.020408} {\bibfield
  {journal} {\bibinfo  {journal} {Phys. Rev. Lett.}\ }\textbf {\bibinfo
  {volume} {109}},\ \bibinfo {pages} {020408} (\bibinfo {year}
  {2012})}\BibitemShut {NoStop}%
\bibitem [{\citenamefont {\ifmmode~\check{C}\else \v{C}\fi{}ernoch}\ \emph
  {et~al.}(2008)\citenamefont {\ifmmode~\check{C}\else \v{C}\fi{}ernoch},
  \citenamefont {Soubusta}, \citenamefont {Bart\ifmmode \mathring{u}\else
  \r{u}\fi{}\ifmmode~\check{s}\else \v{s}\fi{}kov\'a}, \citenamefont
  {Du\ifmmode~\check{s}\else \v{s}\fi{}ek},\ and\ \citenamefont
  {Fiur\'a\ifmmode~\check{s}\else \v{s}\fi{}ek}}]{cernoch2008experimental}%
  \BibitemOpen
  \bibfield  {author} {\bibinfo {author} {\bibfnamefont {A.}~\bibnamefont
  {\ifmmode~\check{C}\else \v{C}\fi{}ernoch}}, \bibinfo {author} {\bibfnamefont
  {J.}~\bibnamefont {Soubusta}}, \bibinfo {author} {\bibfnamefont
  {L.}~\bibnamefont {Bart\ifmmode \mathring{u}\else
  \r{u}\fi{}\ifmmode~\check{s}\else \v{s}\fi{}kov\'a}}, \bibinfo {author}
  {\bibfnamefont {M.}~\bibnamefont {Du\ifmmode~\check{s}\else \v{s}\fi{}ek}}, \
  and\ \bibinfo {author} {\bibfnamefont {J.}~\bibnamefont
  {Fiur\'a\ifmmode~\check{s}\else \v{s}\fi{}ek}},\ }\href {\doibase
  10.1103/PhysRevLett.100.180501} {\bibfield  {journal} {\bibinfo  {journal}
  {Phys. Rev. Lett.}\ }\textbf {\bibinfo {volume} {100}},\ \bibinfo {pages}
  {180501} (\bibinfo {year} {2008})}\BibitemShut {NoStop}%
\bibitem [{Note1()}]{Note1}%
  \BibitemOpen
  \bibinfo {note} {As is usual for a probability, $\protect \mathrm
  {Prob}(x,y)$ is estimated from repeated trials using an identical ensemble of
  input systems. This is implicit for probabilities and expectation values
  throughout the paper.}\BibitemShut {Stop}%
\bibitem [{Note2()}]{Note2}%
  \BibitemOpen
  \bibinfo {note} {The strategy considered here is informationally equivalent
  to separating an identical ensemble into two and measuring $\protect \mathrm
  {Prob}(x)$ with one half and $\protect \mathrm {Prob}(y)$ with the other
  half. Knowing only these two marginal distributions is insufficient to
  determine the quantum state~\cite {lvovsky2009continuous}.}\BibitemShut
  {Stop}%
\bibitem [{\citenamefont {Wootters}\ and\ \citenamefont
  {Zurek}(1982)}]{wootters1982single}%
  \BibitemOpen
  \bibfield  {author} {\bibinfo {author} {\bibfnamefont {W.~K.}\ \bibnamefont
  {Wootters}}\ and\ \bibinfo {author} {\bibfnamefont {W.~H.}\ \bibnamefont
  {Zurek}},\ }\href {\doibase 10.1038/299802a0} {\bibfield  {journal} {\bibinfo
   {journal} {Nature}\ }\textbf {\bibinfo {volume} {299}},\ \bibinfo {pages}
  {802} (\bibinfo {year} {1982})}\BibitemShut {NoStop}%
\bibitem [{\citenamefont {Scarani}\ \emph {et~al.}(2005)\citenamefont
  {Scarani}, \citenamefont {Iblisdir}, \citenamefont {Gisin},\ and\
  \citenamefont {Ac\'{\i}n}}]{scarani2005quantum}%
  \BibitemOpen
  \bibfield  {author} {\bibinfo {author} {\bibfnamefont {V.}~\bibnamefont
  {Scarani}}, \bibinfo {author} {\bibfnamefont {S.}~\bibnamefont {Iblisdir}},
  \bibinfo {author} {\bibfnamefont {N.}~\bibnamefont {Gisin}}, \ and\ \bibinfo
  {author} {\bibfnamefont {A.}~\bibnamefont {Ac\'{\i}n}},\ }\href {\doibase
  10.1103/RevModPhys.77.1225} {\bibfield  {journal} {\bibinfo  {journal} {Rev.
  Mod. Phys.}\ }\textbf {\bibinfo {volume} {77}},\ \bibinfo {pages} {1225}
  (\bibinfo {year} {2005})}\BibitemShut {NoStop}%
\bibitem [{\citenamefont {Bu\ifmmode~\check{z}\else \v{z}\fi{}ek}\ and\
  \citenamefont {Hillery}(1996)}]{buzek1996quantum}%
  \BibitemOpen
  \bibfield  {author} {\bibinfo {author} {\bibfnamefont {V.}~\bibnamefont
  {Bu\ifmmode~\check{z}\else \v{z}\fi{}ek}}\ and\ \bibinfo {author}
  {\bibfnamefont {M.}~\bibnamefont {Hillery}},\ }\href {\doibase
  10.1103/PhysRevA.54.1844} {\bibfield  {journal} {\bibinfo  {journal} {Phys.
  Rev. A}\ }\textbf {\bibinfo {volume} {54}},\ \bibinfo {pages} {1844}
  (\bibinfo {year} {1996})}\BibitemShut {NoStop}%
\bibitem [{\citenamefont {Gisin}\ and\ \citenamefont
  {Massar}(1997)}]{gisin1997optimal}%
  \BibitemOpen
  \bibfield  {author} {\bibinfo {author} {\bibfnamefont {N.}~\bibnamefont
  {Gisin}}\ and\ \bibinfo {author} {\bibfnamefont {S.}~\bibnamefont {Massar}},\
  }\href {\doibase 10.1103/PhysRevLett.79.2153} {\bibfield  {journal} {\bibinfo
   {journal} {Phys. Rev. Lett.}\ }\textbf {\bibinfo {volume} {79}},\ \bibinfo
  {pages} {2153} (\bibinfo {year} {1997})}\BibitemShut {NoStop}%
\bibitem [{\citenamefont {Bruss}\ \emph {et~al.}(1998)\citenamefont {Bruss},
  \citenamefont {Ekert},\ and\ \citenamefont
  {Macchiavello}}]{bruss1998optimal}%
  \BibitemOpen
  \bibfield  {author} {\bibinfo {author} {\bibfnamefont {D.}~\bibnamefont
  {Bruss}}, \bibinfo {author} {\bibfnamefont {A.}~\bibnamefont {Ekert}}, \ and\
  \bibinfo {author} {\bibfnamefont {C.}~\bibnamefont {Macchiavello}},\ }\href
  {\doibase 10.1103/PhysRevLett.81.2598} {\bibfield  {journal} {\bibinfo
  {journal} {Phys. Rev. Lett.}\ }\textbf {\bibinfo {volume} {81}},\ \bibinfo
  {pages} {2598} (\bibinfo {year} {1998})}\BibitemShut {NoStop}%
\bibitem [{\citenamefont {Niset}\ \emph {et~al.}(2007)\citenamefont {Niset},
  \citenamefont {Ac{\'\i}n}, \citenamefont {Andersen}, \citenamefont {Cerf},
  \citenamefont {Garc{\'\i}a-Patr{\'o}n}, \citenamefont {Navascu{\'e}s},\ and\
  \citenamefont {Sabuncu}}]{niset2007superiority}%
  \BibitemOpen
  \bibfield  {author} {\bibinfo {author} {\bibfnamefont {J.}~\bibnamefont
  {Niset}}, \bibinfo {author} {\bibfnamefont {A.}~\bibnamefont {Ac{\'\i}n}},
  \bibinfo {author} {\bibfnamefont {U.~L.}\ \bibnamefont {Andersen}}, \bibinfo
  {author} {\bibfnamefont {N.}~\bibnamefont {Cerf}}, \bibinfo {author}
  {\bibfnamefont {R.}~\bibnamefont {Garc{\'\i}a-Patr{\'o}n}}, \bibinfo {author}
  {\bibfnamefont {M.}~\bibnamefont {Navascu{\'e}s}}, \ and\ \bibinfo {author}
  {\bibfnamefont {M.}~\bibnamefont {Sabuncu}},\ }\href@noop {} {\bibfield
  {journal} {\bibinfo  {journal} {Phys. Rev. Lett.}\ }\textbf {\bibinfo
  {volume} {98}},\ \bibinfo {pages} {260404} (\bibinfo {year}
  {2007})}\BibitemShut {NoStop}%
\bibitem [{\citenamefont {Massar}\ and\ \citenamefont
  {Popescu}(1995)}]{massar1995optimal}%
  \BibitemOpen
  \bibfield  {author} {\bibinfo {author} {\bibfnamefont {S.}~\bibnamefont
  {Massar}}\ and\ \bibinfo {author} {\bibfnamefont {S.}~\bibnamefont
  {Popescu}},\ }\href {\doibase 10.1103/PhysRevLett.74.1259} {\bibfield
  {journal} {\bibinfo  {journal} {Phys. Rev. Lett.}\ }\textbf {\bibinfo
  {volume} {74}},\ \bibinfo {pages} {1259} (\bibinfo {year}
  {1995})}\BibitemShut {NoStop}%
\bibitem [{\citenamefont {D'Ariano}\ \emph {et~al.}(2001)\citenamefont
  {D'Ariano}, \citenamefont {Macchiavello},\ and\ \citenamefont
  {Sacchi}}]{dariano2001joint}%
  \BibitemOpen
  \bibfield  {author} {\bibinfo {author} {\bibfnamefont {G.~M.}\ \bibnamefont
  {D'Ariano}}, \bibinfo {author} {\bibfnamefont {C.}~\bibnamefont
  {Macchiavello}}, \ and\ \bibinfo {author} {\bibfnamefont {M.~F.}\
  \bibnamefont {Sacchi}},\ }\href {http://stacks.iop.org/1464-4266/3/i=2/a=305}
  {\bibfield  {journal} {\bibinfo  {journal} {J. Opt. B: Quantum Semiclass.
  Opt.}\ }\textbf {\bibinfo {volume} {3}},\ \bibinfo {pages} {44} (\bibinfo
  {year} {2001})}\BibitemShut {NoStop}%
\bibitem [{\citenamefont {Brougham}\ \emph {et~al.}(2006)\citenamefont
  {Brougham}, \citenamefont {Andersson},\ and\ \citenamefont
  {Barnett}}]{brougham2006cloning}%
  \BibitemOpen
  \bibfield  {author} {\bibinfo {author} {\bibfnamefont {T.}~\bibnamefont
  {Brougham}}, \bibinfo {author} {\bibfnamefont {E.}~\bibnamefont {Andersson}},
  \ and\ \bibinfo {author} {\bibfnamefont {S.~M.}\ \bibnamefont {Barnett}},\
  }\href {\doibase 10.1103/PhysRevA.73.062319} {\bibfield  {journal} {\bibinfo
  {journal} {Phys. Rev. A}\ }\textbf {\bibinfo {volume} {73}},\ \bibinfo
  {pages} {062319} (\bibinfo {year} {2006})}\BibitemShut {NoStop}%
\bibitem [{SM()}]{SM}%
  \BibitemOpen
  \href@noop {} {}\bibinfo {note} {See supplementary materials for details on
  the experimental setup, derivations, and some additional data. It also
  includes Ref.~[41]}\BibitemShut {NoStop}%
\bibitem [{\citenamefont {Campos}\ and\ \citenamefont
  {Gerry}(2005)}]{campos2005permutation}%
  \BibitemOpen
  \bibfield  {author} {\bibinfo {author} {\bibfnamefont {R.~A.}\ \bibnamefont
  {Campos}}\ and\ \bibinfo {author} {\bibfnamefont {C.~C.}\ \bibnamefont
  {Gerry}},\ }\href {\doibase 10.1103/PhysRevA.72.065803} {\bibfield  {journal}
  {\bibinfo  {journal} {Phys. Rev. A}\ }\textbf {\bibinfo {volume} {72}},\
  \bibinfo {pages} {065803} (\bibinfo {year} {2005})}\BibitemShut {NoStop}%
\bibitem [{\citenamefont {Irvine}\ \emph {et~al.}(2004)\citenamefont {Irvine},
  \citenamefont {Lamas~Linares}, \citenamefont {de~Dood},\ and\ \citenamefont
  {Bouwmeester}}]{irvine2004optimal}%
  \BibitemOpen
  \bibfield  {author} {\bibinfo {author} {\bibfnamefont {W.~T.~M.}\
  \bibnamefont {Irvine}}, \bibinfo {author} {\bibfnamefont {A.}~\bibnamefont
  {Lamas~Linares}}, \bibinfo {author} {\bibfnamefont {M.~J.~A.}\ \bibnamefont
  {de~Dood}}, \ and\ \bibinfo {author} {\bibfnamefont {D.}~\bibnamefont
  {Bouwmeester}},\ }\href {\doibase 10.1103/PhysRevLett.92.047902} {\bibfield
  {journal} {\bibinfo  {journal} {Phys. Rev. Lett.}\ }\textbf {\bibinfo
  {volume} {92}},\ \bibinfo {pages} {047902} (\bibinfo {year}
  {2004})}\BibitemShut {NoStop}%
\bibitem [{\citenamefont {Nagali}\ \emph {et~al.}(2009)\citenamefont {Nagali},
  \citenamefont {Sansoni}, \citenamefont {Sciarrino}, \citenamefont
  {De~Martini}, \citenamefont {Marrucci}, \citenamefont {Piccirillo},
  \citenamefont {Karimi},\ and\ \citenamefont {Santamato}}]{nagali2009optimal}%
  \BibitemOpen
  \bibfield  {author} {\bibinfo {author} {\bibfnamefont {E.}~\bibnamefont
  {Nagali}}, \bibinfo {author} {\bibfnamefont {L.}~\bibnamefont {Sansoni}},
  \bibinfo {author} {\bibfnamefont {F.}~\bibnamefont {Sciarrino}}, \bibinfo
  {author} {\bibfnamefont {F.}~\bibnamefont {De~Martini}}, \bibinfo {author}
  {\bibfnamefont {L.}~\bibnamefont {Marrucci}}, \bibinfo {author}
  {\bibfnamefont {B.}~\bibnamefont {Piccirillo}}, \bibinfo {author}
  {\bibfnamefont {E.}~\bibnamefont {Karimi}}, \ and\ \bibinfo {author}
  {\bibfnamefont {E.}~\bibnamefont {Santamato}},\ }\href {\doibase
  10.1038/nphoton.2009.214} {\bibfield  {journal} {\bibinfo  {journal} {Nat.
  Photon.}\ }\textbf {\bibinfo {volume} {3}},\ \bibinfo {pages} {720} (\bibinfo
  {year} {2009})}\BibitemShut {NoStop}%
\bibitem [{\citenamefont {Bouchard}\ \emph {et~al.}(2017)\citenamefont
  {Bouchard}, \citenamefont {Fickler}, \citenamefont {Boyd},\ and\
  \citenamefont {Karimi}}]{bouchard2016high}%
  \BibitemOpen
  \bibfield  {author} {\bibinfo {author} {\bibfnamefont {F.}~\bibnamefont
  {Bouchard}}, \bibinfo {author} {\bibfnamefont {R.}~\bibnamefont {Fickler}},
  \bibinfo {author} {\bibfnamefont {R.~W.}\ \bibnamefont {Boyd}}, \ and\
  \bibinfo {author} {\bibfnamefont {E.}~\bibnamefont {Karimi}},\ }\href
  {\doibase 10.1126/sciadv.1601915} {\bibfield  {journal} {\bibinfo  {journal}
  {Science Adv.}\ }\textbf {\bibinfo {volume} {3}} (\bibinfo {year} {2017}),\
  10.1126/sciadv.1601915}\BibitemShut {NoStop}%
\bibitem [{\citenamefont {MacLean}\ \emph {et~al.}(2016)\citenamefont
  {MacLean}, \citenamefont {Ried}, \citenamefont {Spekkens},\ and\
  \citenamefont {Resch}}]{maclean2016quantum}%
  \BibitemOpen
  \bibfield  {author} {\bibinfo {author} {\bibfnamefont {J.-P.~W.}\
  \bibnamefont {MacLean}}, \bibinfo {author} {\bibfnamefont {K.}~\bibnamefont
  {Ried}}, \bibinfo {author} {\bibfnamefont {R.~W.}\ \bibnamefont {Spekkens}},
  \ and\ \bibinfo {author} {\bibfnamefont {K.~J.}\ \bibnamefont {Resch}},\
  }\href@noop {} {\enquote {\bibinfo {title} {Quantum-coherent mixtures of
  causal relations},}\ } (\bibinfo {year} {2016}),\ \Eprint
  {http://arxiv.org/abs/arXiv:1606.04523} {arXiv:1606.04523} \BibitemShut
  {NoStop}%
\bibitem [{\citenamefont {Andersen}\ \emph {et~al.}(2005)\citenamefont
  {Andersen}, \citenamefont {Josse},\ and\ \citenamefont
  {Leuchs}}]{andersen2005unconditional}%
  \BibitemOpen
  \bibfield  {author} {\bibinfo {author} {\bibfnamefont {U.~L.}\ \bibnamefont
  {Andersen}}, \bibinfo {author} {\bibfnamefont {V.}~\bibnamefont {Josse}}, \
  and\ \bibinfo {author} {\bibfnamefont {G.}~\bibnamefont {Leuchs}},\
  }\href@noop {} {\bibfield  {journal} {\bibinfo  {journal} {Phys. Rev. Lett.}\
  }\textbf {\bibinfo {volume} {94}},\ \bibinfo {pages} {240503} (\bibinfo
  {year} {2005})}\BibitemShut {NoStop}%
\bibitem [{\citenamefont {Sabuncu}\ \emph {et~al.}(2007)\citenamefont
  {Sabuncu}, \citenamefont {Andersen},\ and\ \citenamefont
  {Leuchs}}]{sabuncu2007experimental}%
  \BibitemOpen
  \bibfield  {author} {\bibinfo {author} {\bibfnamefont {M.}~\bibnamefont
  {Sabuncu}}, \bibinfo {author} {\bibfnamefont {U.~L.}\ \bibnamefont
  {Andersen}}, \ and\ \bibinfo {author} {\bibfnamefont {G.}~\bibnamefont
  {Leuchs}},\ }\href@noop {} {\bibfield  {journal} {\bibinfo  {journal} {Phys.
  Rev. Lett.}\ }\textbf {\bibinfo {volume} {98}},\ \bibinfo {pages} {170503}
  (\bibinfo {year} {2007})}\BibitemShut {NoStop}%
\bibitem [{\citenamefont {Patel}\ \emph {et~al.}(2016)\citenamefont {Patel},
  \citenamefont {Ho}, \citenamefont {Ferreyrol}, \citenamefont {Ralph},\ and\
  \citenamefont {Pryde}}]{patel2016quantum}%
  \BibitemOpen
  \bibfield  {author} {\bibinfo {author} {\bibfnamefont {R.~B.}\ \bibnamefont
  {Patel}}, \bibinfo {author} {\bibfnamefont {J.}~\bibnamefont {Ho}}, \bibinfo
  {author} {\bibfnamefont {F.}~\bibnamefont {Ferreyrol}}, \bibinfo {author}
  {\bibfnamefont {T.~C.}\ \bibnamefont {Ralph}}, \ and\ \bibinfo {author}
  {\bibfnamefont {G.~J.}\ \bibnamefont {Pryde}},\ }\href {\doibase
  10.1126/sciadv.1501531} {\bibfield  {journal} {\bibinfo  {journal} {Science
  Adv.}\ }\textbf {\bibinfo {volume} {2}} (\bibinfo {year} {2016}),\
  10.1126/sciadv.1501531}\BibitemShut {NoStop}%
\bibitem [{\citenamefont {Durt}\ \emph {et~al.}(2010)\citenamefont {Durt},
  \citenamefont {Englert}, \citenamefont {Bengtsson},\ and\ \citenamefont
  {{\.Z}yczkowski}}]{durt2010mutually}%
  \BibitemOpen
  \bibfield  {author} {\bibinfo {author} {\bibfnamefont {T.}~\bibnamefont
  {Durt}}, \bibinfo {author} {\bibfnamefont {B.-G.}\ \bibnamefont {Englert}},
  \bibinfo {author} {\bibfnamefont {I.}~\bibnamefont {Bengtsson}}, \ and\
  \bibinfo {author} {\bibfnamefont {K.}~\bibnamefont {{\.Z}yczkowski}},\
  }\href@noop {} {\bibfield  {journal} {\bibinfo  {journal} {Int. J. Quantum
  Inform.}\ }\textbf {\bibinfo {volume} {8}},\ \bibinfo {pages} {535} (\bibinfo
  {year} {2010})}\BibitemShut {NoStop}%
\bibitem [{\citenamefont {Lvovsky}\ and\ \citenamefont
  {Raymer}(2009)}]{lvovsky2009continuous}%
  \BibitemOpen
  \bibfield  {author} {\bibinfo {author} {\bibfnamefont {A.~I.}\ \bibnamefont
  {Lvovsky}}\ and\ \bibinfo {author} {\bibfnamefont {M.~G.}\ \bibnamefont
  {Raymer}},\ }\href {\doibase 10.1103/RevModPhys.81.299} {\bibfield  {journal}
  {\bibinfo  {journal} {Rev. Mod. Phys.}\ }\textbf {\bibinfo {volume} {81}},\
  \bibinfo {pages} {299} (\bibinfo {year} {2009})}\BibitemShut {NoStop}%
\end{thebibliography}%

\newpage
\onecolumngrid

\section*{Supplementary Material}

\renewcommand{\theequation}{S\arabic{equation}}
\renewcommand{\thefigure}{S\arabic{figure}}

\section*{Experimental Setup}
A detailed figure containing the experimental setup is shown in Fig.~S1. A 40 mW continuous-wave diode laser at 404 nm pumps a type-II $\beta$-barium borate crystal. Through spontaneous parametric down-conversion, pairs of 808 nm photons with orthogonal polarization are generated collinearly with the pump laser. The latter is then blocked by a long pass filter. The photon pair splits at a polarizing beam splitter (PBS), and each photon is coupled into a polarization-maintaining single mode fiber. The path length difference between the photon paths is adjusted with a delay stage. A spinning (2 Hz) half-wave plate produces a completely mixed state $\bm{I}/2$ at one fiber output, while a half-wave plate and quarter-wave plate produce the state $\bm{\rho}$ to be cloned at the other fiber output. A displaced Sagnac interferometer composed of two BS is used instead of the interferometer in Fig.~1 (of main text), since it is more robust to air fluctuations and other instabilities. The phase $\varphi$ between red and blue paths is adjusted by slightly rotating one of the mirrors in the interferometer in order to change the path length difference between both paths. A series of wave plates and a PBS are used to implement the projectors $\bm{x}$ and $\bm{y}$. Detectors are single photon counting silicon avalanche photodiodes. Using time-correlation electronics, we count coincidence events that occur in a 5 nanosecond window and average over 60 seconds for each measurement. 

\section*{Joint measurement on optimal clones}
For qudits, the $d$-dimensional observables $\bm{X}$ and $\bm{Y}$ are complementary if their eigenstates $\{ \ket{x}\}$ and $\{\ket{y}\}$ all satisfy $\left| \braket{x|y} \right| = 1/\sqrt{d}$. We use the notation $\bm{x} = \ket{x}\bra{x}$. The output of an optimal cloner for qudits is:
\begin{equation}
\bm{o}_{ab}^j = \frac{2}{d+1}\left( \bm{\Pi}^{j}_{ab} \bm{\rho}_a \bm{I}_b \bm{\Pi}^{j\dagger}_{ab} \right)
\end{equation}
with $j=+1$. Consider measuring $\bm{X}$ in mode $a$ and $\bm{Y}$ in mode $b$. As shown in Ref.~\cite{hofmann2012how}, the joint probability of measuring outcome $X=x$ and $Y=y$ is:
\begin{equation}
\begin{aligned}
\label{eqn:opt_clone_jm}
\mathrm{Prob}^j(x,y) &= \mathrm{Tr}\left[\bm{x}_a \bm{y}_b \bm{o}_{ab}^j\right] \\
&= \frac{1}{2(d+1)}\left( \braket{\bm{x}}_{\bm{\rho}} + \braket{\bm{y}}_{\bm{\rho}} + 2\mathrm{Re}\left (j\braket{\bm{xy}}_{\bm{\rho}}\right ) \right).
\end{aligned}
\end{equation}
The terms $\braket{\bm{x}}_{\bm{\rho}}$ and $\braket{\bm{y}}_{\bm{\rho}}$ could be obtained from a joint measurement on trivial clones. In contrast, the last term $\braket{\bm{xy}}_{\bm{\rho}}$ is obtained from a joint measurement on twins, and is a joint quasiprobability of simultaneously measuring both $\bm{x}$ and $\bm{y}$ on $\bm{\rho}$. In order to isolate the latter term, we use the fact that the joint measurement contribution of the trivial clones does not depend on the phase $j$, giving:
\begin{equation}
\label{eqn:weak_average}
\braket{\bm{xy}}_{\bm{\rho}}  = \frac{d+1}{2} \sum_{j = \pm 1, \pm i} j^* \mathrm{Prob}^j(x,y).
\end{equation}

When the input state is pure, that is $\bm{\rho} = \ket{\psi}\bra{\psi}$, then $\braket{\bm{xy}}_{\psi} = \nu\braket{y|\psi}$, where $\nu = \braket{\psi|x}\braket{x|y}$. For some $x=x_0$, the phase of $\nu$ is constant for all $y$ and so the wave function $\ket{\psi}$ can be expressed in the basis of $\bm{Y}$ as $\ket{\psi} = \frac{1}{\nu}\sum_y\braket{\bm{x_0y}}_\psi \ket{y}$. As usual, the constant $\nu$ is found by normalizing $\ket{\psi}$. Thus, using Eq.~\ref{eqn:weak_average}, any complex amplitude $\psi(y)  = \braket{y|\psi}$ of the wave function can be found from:
\begin{equation}
 \psi(y) = \frac{d+1}{2\nu} \sum_{j = \pm 1, \pm i} j^* \mathrm{Prob}^j(x_0,y).
\end{equation}
The choice of $x_0$ is equivalent to choosing a phase reference for the wave function. In Fig.~3A of the main text, we use $x_0 = d$, which defines the diagonal polarization as $\ket{d} = (\ket{h} + \ket{v})/\sqrt{2}$. We choose to make the normalization constant $\nu$ a real number, \textit{i.e.} $\nu = ( \left|\braket{\bm{dh}}\right|^2 + \left|\braket{\bm{dv}}\right|^2 )^{1/2}$.

For mixed input states, the joint quasiprobability $\braket{\bm{xy}}_{\bm{\rho}}$ is related to the density matrix $\bm{\rho}$ via a discrete Fourier transform (see derivation below).

\section*{Relating the joint quasiprobability to the density matrix}

Here we summarize the connection between the joint quasiprobability distribution and the density matrix. Consider the $d$-dimensional complementary observables $\bm{X}$ and $\bm{Y}$ with eigenstates $\{ x_i \}$ and $\{ y_i \}$ such that $\left|\braket{x_i|y_j}\right|=1/\sqrt{d}$ for any $i,j$. Without loss of generality~\cite{durt2010mutually}, one can take the $\{ y_i \}$ basis to be defined in terms of a discrete Fourier transform of $\{ x_i \}$: $\ket{y_j} = \sum_{i=0}^{d-1} \ket{x_i}\exp{(i2\pi x_i y_j/d)}/\sqrt{d}$. This fixes a phase relation for the inner product of all the eigenstates of both bases:
\begin{equation}
\braket{x_i|y_j} = \exp{(i2\pi x_i y_j)} / \sqrt{d}.
\end{equation}
A general $d$-dimensional quantum state $\bm{\rho}$ can be written in the basis of $\bm{Y}$ as $\bm{\rho} = \sum_{k,l} p_{kl} \ket{y_k}\bra{y_l}$. We wish to relate the coefficients $p_{kl}$ to the joint quasiprobability distribution described in the main text, \textit{i.e.} $\braket{\bm{xy}}_{\bm{\rho}}$. Recall that we use the notation $\bm{x} = \ket{x}\bra{x}$. Thus $\braket{\bm{xy}}_{\bm{\rho}}$ takes the form of $D_{ij} \equiv \braket{\ket{x_i}\braket{x_i|y_j}\bra{y_j}}_{\bm{\rho}} = \mathrm{Tr}\left[\ket{x_i}\braket{x_i|y_j}\bra{y_j}\bm{\rho}\right] = \braket{x_i|y_j}\braket{y_j|\bm{\rho}|x_i}$. Inserting the expanded form of $\bm{\rho}$:
\begin{equation}
D_{ij} = \braket{x_i|y_j}\braket{y_j| \sum_{k,l}p_{kl}|y_k}\braket{y_l|x_i} = \braket{x_i|y_j} \sum_l p_{jl} \braket{y_l|x_i} =  \frac{\exp{(i2\pi x_i y_j)}}{d}\sum_{l=0}^{d-1}p_{jl}\exp{(-i2\pi x_i y_l)}.
\end{equation}
This shows that the joint quasiprobability is the discrete Fourier transform of the density matrix. The equation can be inverted by taking the inverse Fourier transform of both sides: 
\begin{equation}
\label{eqn:wa_to_dm}
p_{jl} = \sum_{i=0}^{d-1} D_{ij} \exp{(i2\pi x_i (y_l-y_j))}.
\end{equation}

In the case of polarization qubits, $\bm{x} = \{ \bm{d},\bm{a} \}$ and $\bm{y} = \{ \bm{h},\bm{v} \}$. Then the equation relating the two density matrix and the joint quasiprobability is:
\begin{equation}
\label{eqn:density_matrix}
\bm{\rho} = \begin{pmatrix} 
\braket{\bm{dh}}_{\bm{\rho}} + \braket{\bm{ah}}_{\bm{\rho}} & \braket{\bm{dh}}_{\bm{\rho}} - \braket{\bm{ah}}_{\bm{\rho}} \\
\braket{\bm{dv}}_{\bm{\rho}} - \braket{\bm{av}}_{\bm{\rho}} &
\braket{\bm{dv}}_{\bm{\rho}} + \braket{\bm{av}}_{\bm{\rho}}
\end{pmatrix}.
\end{equation}
Eq.~\ref{eqn:density_matrix} is used to calculate the density matrix in Fig.~3C of the main text.

\section*{Trivial to optimal clones}
When the two input photons are temporally distinguishable, Hong-Ou-Mandel interference does not occur at the first beam splitter, and the cloner produces trivial clones $\bm{t}_{ab}$. Conversely, for temporally indistinguishable photons, we produce $\bm{o}^j_{ab}$. Photons that are partially distinguishable can be decomposed into the form 
\begin{equation}
\bm{\sigma}^j_{ab} = |\alpha|^2\bm{o}^j_{ab} + (1-|\alpha|^2)\bm{t}_{ab},
\end{equation}
where $\alpha \in [0,1]$ is a temporal distinguishability factor. In particular, the temporal mode of the delayed photon in mode $a$ can be written as $\ket{\zeta_a} = \int d\omega \phi(\omega)e^{-i\omega \tau}a^\dagger(\omega)\ket{0}$ where $\tau$ is the delay, while the other photon in mode $b$ is described by $\ket{\zeta_b} = \int d\omega \phi(\omega)b^\dagger(\omega)\ket{0}$. For a Gaussian spectral amplitude $\phi(\omega) = \frac{1}{\sqrt{\pi}\Delta\omega}e^{\frac{-(\omega - \omega_0)^2}{2\Delta\omega^2}}$ where $\omega_0$ is the central frequency of the photons and $\Delta\omega$ is their spectral width, the distinguishability factor is given by 
\begin{equation}
|\alpha|^2 = \left|\braket{\zeta_a|\zeta_b}\right|^2 = e^{\frac{-\Delta\omega^2\tau^2}{2}}. 
\end{equation}
In the experiment, we adjust the delay $\tau$ by moving the delay stage. The parameter $\Delta\omega^2$ is extracted from fitting a Gaussian to the Hong-Ou-Mandel dip (see Fig. S2).

\section*{Extended Data}

\begin{figure*}[h!]
\centering
\includegraphics[width=0.8\textwidth]{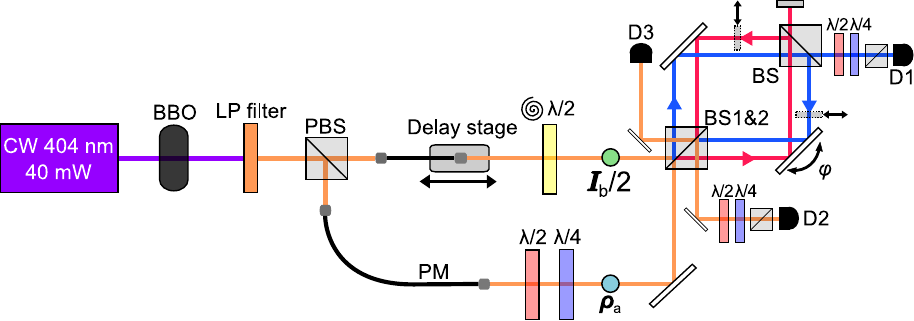}
\caption{\textbf{Experimental setup.} Details of the experimental setup can be found in the Methods section. A simplified schematic of this setup is shown in the main text. Detector D3 is used for alignment purposes, but otherwise is not used in the experiment. CW: continuous-wave, BBO: $\beta$-barium borate, LP: long pass, (P)BS: (polarizing) beam splitter, PM: polarization-maintaining, $\lambda / 2$: half-wave plate, $\lambda / 4$: quarter-wave plate, D: avalanche photodiode detector.}
\label{fig:exp_setup}
\end{figure*}

\begin{figure*}[h!]
\centering
\includegraphics[width=0.5\textwidth]{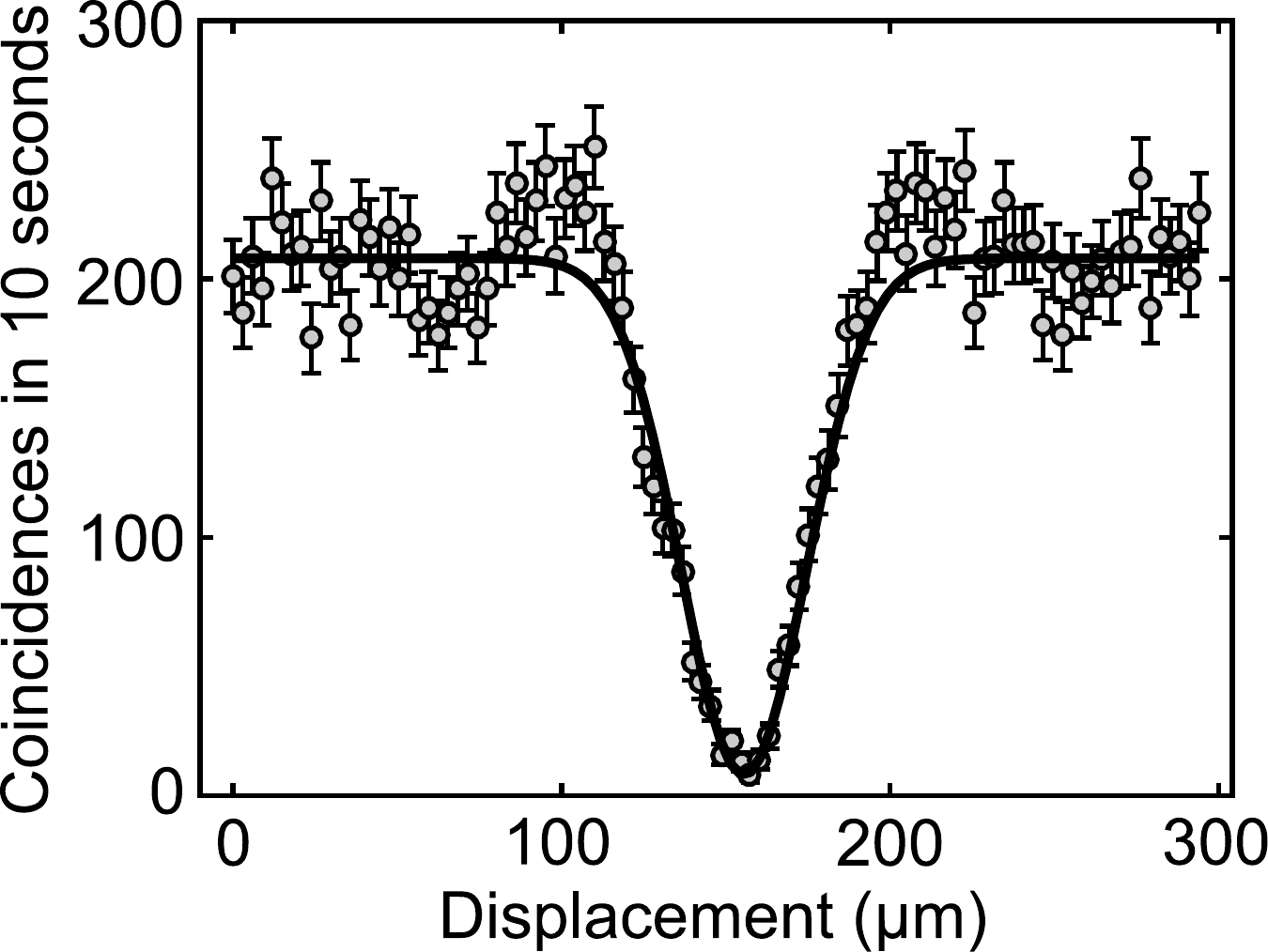}
\caption{\textbf{Hong-Ou-Mandel interference.} In order to characterize the spectral width of the photons and to ensure that we are performing the symmetry projector $\bm{\Pi}^{\pm 1}_{ab}$, we measure the width and visibility of the Hong-Ou-Mandel dip at BS1. With both input photons horizontally polarized and the blue path blocked, we measure the number of coincidences at detectors D1 and D2 as a function of the position of the delay stage. The visibility $\mathcal{V} = (C_{max} - C_{min})/(C_{max} + C_{min})$ (where $C$ is the number of coincidences) of the dip is $\sim$ 96\%. Error bars are calculated using Poissonian counting statistics.}
\label{fig:dip}
\end{figure*}

\begin{figure*}[h!]
\centering
\includegraphics[width=0.8\textwidth]{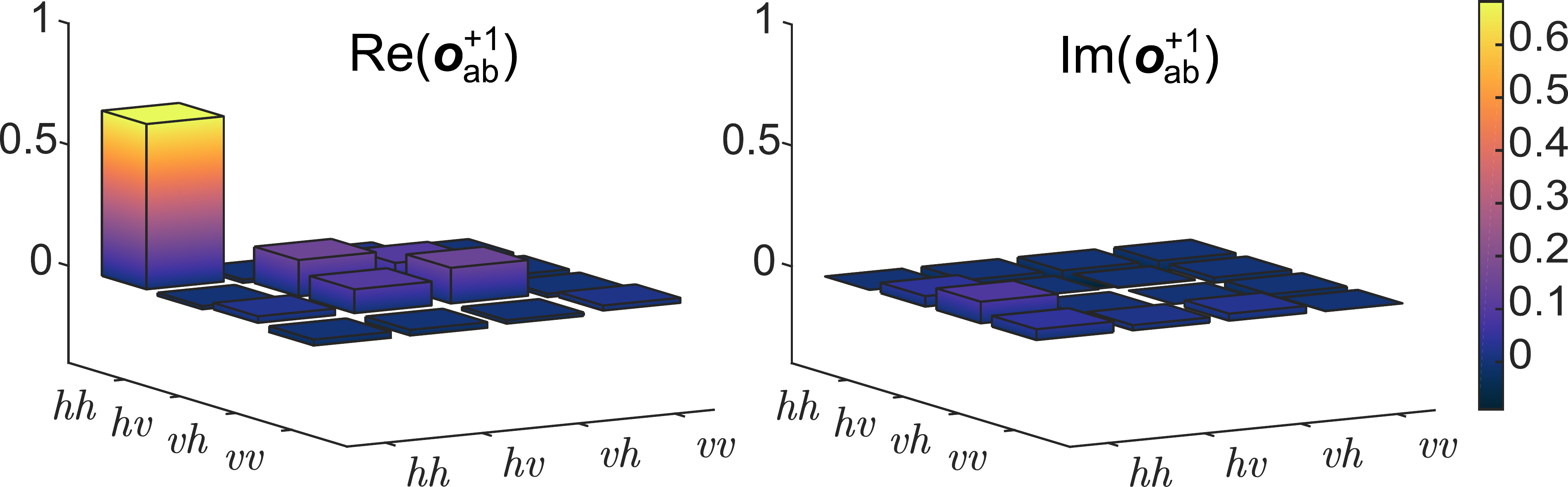}
\caption{\textbf{Quantum state tomography of $\bm{\Pi}^{+1}_{ab}$ output.} In order to determine the fidelity of our clones, we perform two-photon quantum state tomography on the output $\bm{o}^{+1}_{ab}$ of the cloner. Here the input state to be cloned is $\bm{\rho}_a = \bm{h}$. By tracing over each subsystem of the measured $\bm{o}^{+1}_{ab}$, we can compute the fidelities  $ F_a = \left|\braket{h|\bm{o}^{+1}_a|h}\right|^2$ and $ F_b = \left| \braket{h|\bm{o}^{+1}_b|h}\right|^2$. We obtain $F_a = 0.832$ and $F_b = 0.829$, which nearly saturates the theoretical bound of $5/6\sim 0.833$.}
\label{fig:clone}
\end{figure*}

\begin{figure*}[h!]
\centering
\includegraphics[width=0.8\textwidth]{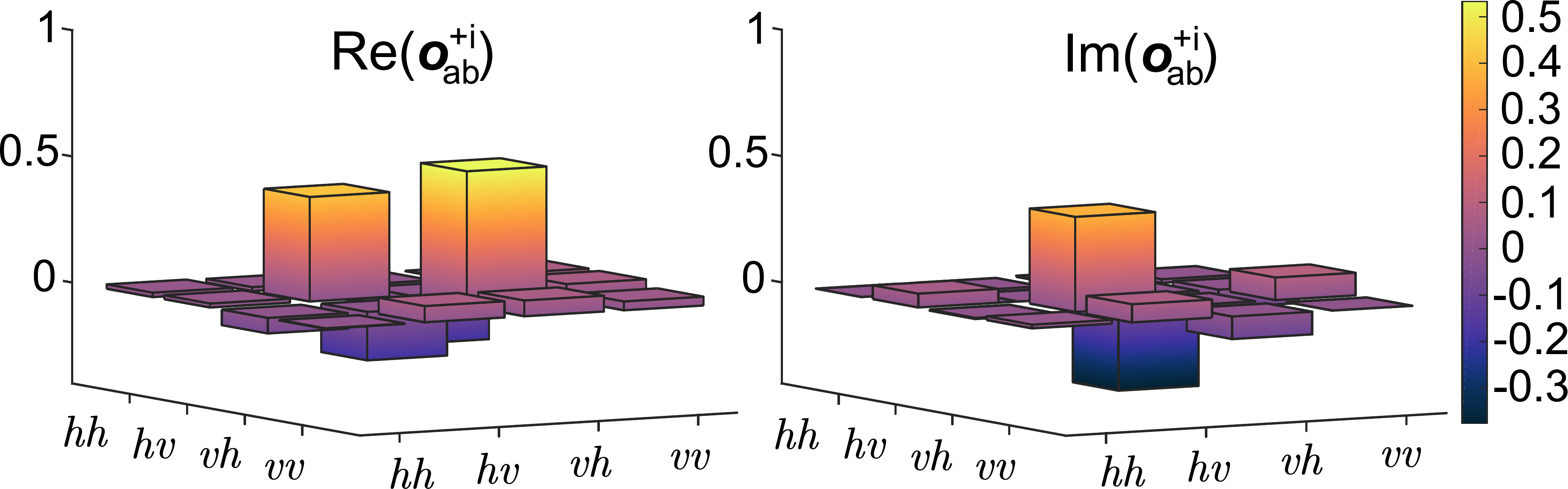}
\caption{\textbf{Quantum state tomography of $\bm{\Pi}_{ab}^{+i}$ output.} In order to achieve the $\bm{\Pi}^{+i}_{ab}$ operation, both paths in the interferometer are unblocked and the phase between them is $\varphi = \pi/2$. We test our ability to implement $\bm{\Pi}^{+i}_{ab}$ by performing two-photon quantum state tomography on the state after the $\bm{\Pi}^{+i}_{ab}$ operation. As an input, we use the state $\ket{hv}\bra{hv}$. In this case, the fidelity of the output state $\bm{o}^{+i}_{ab}$ is 0.850.}
\label{fig:sqrtswap}
\end{figure*}

\begin{figure*}[h!]
\centering
\includegraphics[width=0.5\textwidth]{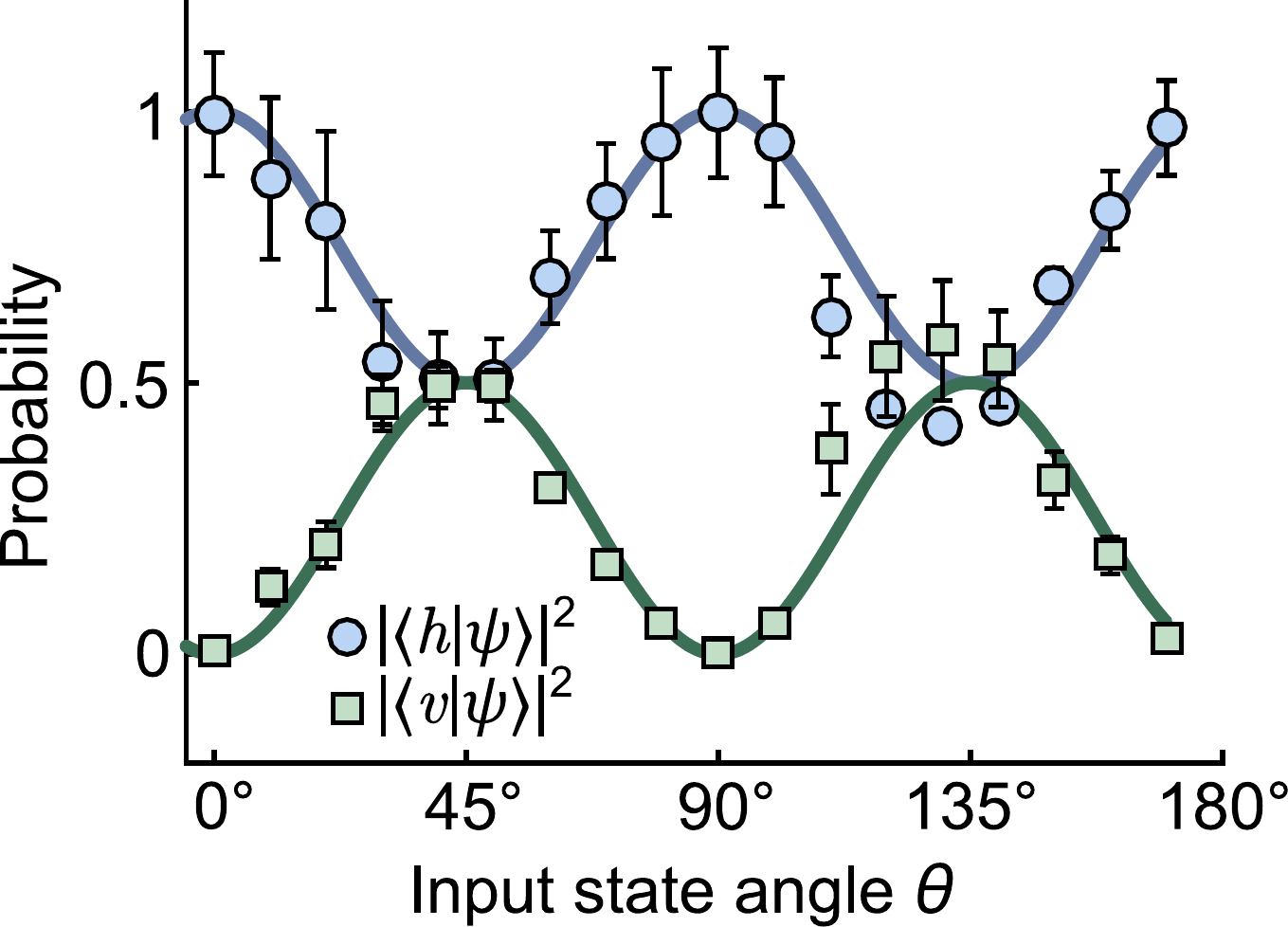}
\caption{\textbf{Absolute value squared of the measured wave function.} The data in this figure is the same as the data used in Fig.~3 of the main text. The polarization state of the input photon as a function of the the quarter wave-plate fast-axis can be written in the form $\ket{\psi} = \alpha\ket{h} + \beta\ket{v}$. Here we plot both $|\alpha|^2 = \cos^4{\theta} + \sin^4{\theta}$ and $|\beta|^2 = 2\sin^2{\theta}\cos^2{\theta}$ (theory is bold lines). Error bars are calculated using Poissonian counting statistics.}
\label{fig:sqrtswap}
\end{figure*}

\end{document}